# Radiation Eigenmodes of Dicke Superradiance

Benjamin Lemberger, Klaus Mølmer

Center for Complex Quantum Systems, Department of Physics and Astronomy, Aarhus University, Ny Munkegade 120, DK-8000 Aarhus C, Denmark.

**ABSTRACT**

We calculate the field eigenmodes of the superradiant emission from an ensemble of $N$ two-level atoms. While numerical techniques are effective due to the symmetry of the problem, we develop also an analytical method to approximates the modes in the limit of a large number of emitters. We find that Dicke superradiant emission is restricted to a small number of modes, with a little over 90% of the photons emitted in a single dominant mode.

## 1. Introduction

Superradiance is a phenomenon that occurs in ensembles of excited quantum systems coupled to a common bosonic bath, for example a cloud of atoms undergoing spontaneous emission. The atomic dynamics of such a system, restricted to spatial separations well below the wavelength of the emitted radiation, were studied by Dicke in his original paper [1]. The intensity of the outgoing radiation can be deduced by the rate of change of the energy contained in the atomic system. Unlike single atom exponential decay, the maximal intensity in superradiance is not at $t=0$ when there is the largest excitation of the atoms, but rather when approximately half of the atoms are excited. A finite time after initial maximal excitation of the atoms a short burst of radiation is emitted with a peak intensity growing as $N^2$, rather than the expected $N$ for independently emitting atoms.

Superradiance is relevant to new laser mechanisms [2,3], matter wave collective dynamics [4], quantum memories [5], quantum computers [6,7] and quantum metrology [8]. Since Dicke's work theoretical research in superradiance [9–11] has been extended to account, e.g., for more complex atomic structures and finite sample sizes, and experimental signatures of superradiance have been observed in a wide variety of physical systems including radio [12] and microwave [13] frequency fields, quantum dots [14], and coupling of quantum emitters to waveguides and complex photonic structures [15–18]. In larger samples, the complimentary collective effect of subradiance has also been analyzed for single and multiple excitations [16,19].

The essential properties of superradiance can be observed in the simplified model used by Dicke, in which the atoms are contained in an infinitesimal volume and modeled as simple two level quantum systems (TLS). The symmetry of this model under the permutation of atoms constrains the Hilbert space to be effectively $N+1$ dimensional instead of $2^N$ dimensional and permits numerical and analytic approaches which are not applicable in general.

The primary focus of early studies of superradiance was on the time dependent intensity profile, while the statistical properties of the emitted superradiance signal were addressed by Glauber and Haake [20,21], who used the simple linear relation between the field operators and the atomic dipole operators, see e.g., [22], to calculate first and higher moments of the intensity emitted by an extended ensemble of two-level emitters. They also showed that a semiclassical stochastic model with individual superradiant pulses is able to quantitatively reproduce the intensity fluctuations. Each of these pulses describe a coherent state occupying the specified temporal mode as defined by Titualer and Glauber [23]. While the stochastic pulses may attain a vast number of shapes, the Karhunen-Loeve Theorem ensures that the two-time correlation function of the stochastic field signal, can be compactly expressed as an expansion over a discrete set of modes, $\langle E^*(t)E(t')\rangle = \sum_i n_i v_i^*(t) v_i(t')$.

The Karhunen-Loeve expansion also applies for the characterization of non-classical states of the quantized field, where it allows a modal expansion of the first order coherence function of the time dependent quantized field operators, evaluated in the Heisenberg picture. Note that the modes identified serve as classical wave packet solutions on which the quantized field is expanded with appropriate annihilation and creation operators. Raymer et al [24] thus used this description to provide a few mode expansion of the quantum fluctuations in stimulated Raman scattering. For a recent review article on the early works on expansion of quantized fields on physically motivated sets of temporal modes, we refer to [25] , see also [26], [27].

Precise knowledge of the outgoing modes is particularly relevant to applications in quantum information and quantum metrology protocols that often rely on single mode assumptions. In this article we identify the most populated field modes of superradiant emission in the Dicke model. We will find that, in the large $N$ limit, the majority ( $>$ 90 %) of outgoing photons are contained in a single mode, and the remaining $\sim 10$ % are mostly contained in a small number of other modes.

The article is organized as follows; in Section 2 we define the model of the atomic system at hand and write down the differential equations obeyed by its state components. We show how to derive the outgoing electric field modes from the atomic dynamics and numerically solve those dynamics to find the modes for $N = 60$. In Section 3 we develop an approximate analytic solution to the differential equations presented in Section 2, and use it to find a closed form approximation for the atomic two-time correlation function. We can diagonalize this function to again find the electric modes, as well as analytic expressions for the emitted field intensity. Section 4 contains our conclusions, and in the Appendix we extend the analytic approximation to find a simple set of differential equations which can be intergrated iteratively to find the outgoing field modes in (descending) order of occupation.

## 2. Atomic System

Consider an ensemble of $N$ two level systems with energy difference $\hbar\omega$ between the ground and excited states, and suppose that the length of the entire sample, in each dimension, is much less than the wavelength of light associated with the transition frequency $\omega$. We assume that the TLSs do not interact with each other directly but are coupled to the quantized electromagnetic field, taken to be a bosonic bath at zero temperature. We consider an initial state where all TLSs are excited. The permutation symmetry of the system constrains the dynamics to a Hillbert space of dimension $N+1$, corresponding to each possible total number of excitations $m$ shared by the ensemble.

We will study the properties of the outgoing electromagnetic field through use of the master equation and the quantum regression theorem.

## 2.1. Dicke dynamics

By application of the Born Markov approximation to the Schrödinger Equation for the atoms and electromagnetic field in free space [28], we eliminate the bosonic bath degrees of freedom and obtain a master equation for the reduced density matrix of the emitters,

$$\dot{\rho} = \Gamma\big(-\{J^+J^-, \rho\}_+ + J^-\rho J^+\big), \qquad (1)$$

$\Gamma$ is the spontaneous emission rate for a single emitter and energy shifts due to the interaction with the bath are neglected, or incorporated in the definition of the transition frequency $\omega$. In (1), $J^+$ and $J^-$ are the collective raising and lowering operators,

$$J^\pm = \sum_{i_1}^N J_i^\pm \qquad (2)$$

where $J_i^{+(-)}$ is the raising (lowering) operator for the $i$th dipole emitter.

The master equation (1) applies for very compact samples of emitters and for larger samples where the emitters are located at the field anti-nodes of a cavity standing wave, such that they all couple to the cavity field mode with the same strength $g$. If the cavity field is damped with the rate $\kappa$, this leads to the Purcell effect where the emitters undergo collective superradiant emission described by (1) with $\Gamma = 2g^2/\kappa$ [29].

From the definition of the single raising and lowering operators, we obtain their action on the symmetric states $|m\rangle$ with $m$ excited emitters [1,9],

$$\begin{aligned} J^- |m\rangle &= \sqrt{m(N-m+1)}\,|m-1\rangle \\ J^+ |m-1\rangle &= \sqrt{m(N-m+1)}\,|m\rangle. \end{aligned} \qquad (3)$$

The population of the state with $m$ excitations decays with the rate $\Gamma_m = \Gamma m(N - m + 1)$, and (1) yields the coupled equations for the density matrix elements,

$$\dot{\rho}_{mm'} = -\frac{1}{2}(\Gamma_m + \Gamma_{m'})\rho_{mm'} + \sqrt{\Gamma_{m+1}\Gamma_{m'+1}}\,\rho_{m+1,m'+1}. \qquad (4)$$

With the initial condition $\rho(0)_{NN} = 1$ while all other elements vanish, $\rho(t)$ will only develop non-zero values along the diagonal, and we get a closed set of equations for the diagonal elements $\pi_m = \rho_{mm}$:

$$\dot{\pi}_m = -\Gamma_m \pi_m + \Gamma_{m+1}\pi_{m+1}. \qquad (5)$$

## 2.2. The emitted field

The quantized electrolmagnetic field emitted by the excited atoms can be expressed in the Heisenberg picture in terms of the incoming vacuum field operators and the dipole operators of the atoms [22]. This permits evaluation of mean fields and intensities. We are interested in the two-time correlation function

$$\langle E_-(t')E_+(t)\rangle \tag{6}$$

where $E_{-(+)}(t)$ engages the creation (annihilation) operator of the field at the location of detection. Also for the calculation of the field correlation function, we can express the field operators in terms of the atomic dipole raising and lowering operators, i.e., $\langle E_-(t')E_+(t)\rangle \propto \Gamma\langle J^+(t')J^-(t)\rangle$. The proportionality sign applies if only part of the signal is recorded, but for the compact sample, the emission occurs into an angular dipole mode, and for the emission mediated by a single cavity mode, the emitted field occupies a single transverse mode, and our aim in the following is to identify the temporal, i.e., radial or longitudinal, dependence of the most occupied modes.

Like in the case of classical noisy fields, these modes can be identified by the Karhunen-Loeve expansion,

$$\Gamma\langle J^+(t')J^-(t)\rangle = \sum_i n_i v_i(t')^* v_i(t) \tag{7}$$

where $n_i$ denotes the expectation value of the number of photons of each mode $v_i(t)$. If we know the correlation function on a discrete time grid, the expansion (7) is analogous to the eigenvalue decomposition of a finite matrix while the continuous case follows by Mercer's theorem [30]. We note that in contrast to the determination of steady state fluorescence spectra which only depend, by the Fourier transform, on the time difference in the temporal correlation functions [22,31], our mode expansion is obtained for a transient phenomenon, and the dependence of the correlation function on two time arguments is crucial to determine the modes.

The two-time correlation function can be evaluated by the master equation and the quantum regression theorem [28]. For $t < t'$, we must first evaluate $\rho(t)$ by solving the master equation (1) until time $t$. We then multiply the operator $J^-$ on $\rho(t)$ and treat the resulting matrix as initial state for further evolution by the same equation (1) until time $t'$. Finally we multiply by $J^+$ and evaluate the trace.

In practice, we thus solve the linear set of equations (5) until time $t$. Inserting the values of $\pi_m(t)$ as diagonal elements in a matrix and multiplying with the operator $J^-$ yields a matrix with the only non-vanishing elements along the subdiagonal. We denote the $m, m' = m+1$ element of this matrix by $\xi_m$, and by the Quantum Regression Theorem, we obtain the $N$ coupled equations,

$$\dot{\xi}_m = -\frac{1}{2}(\Gamma_m + \Gamma_{m+1})\xi_m + \sqrt{\Gamma_{m+1}\Gamma_{m+2}}\xi_{m+1} \tag{8}$$

from (1).

Eqs.(8) have to be solved from $t$ until $t'$ with the initial condition, $\xi_m(t) = \sqrt{\Gamma}(J^-\rho(t))_{m,m+1} = \sqrt{\Gamma}\sqrt{(m+1)(N-m)}\pi_m(t)$.

After the evolution of these equations from $t$ to $t'$, noting that they represent the non-vanishing lower diagonal elements of a matrix, the multiplication of this matrix by $\sqrt{\Gamma}J^+$ leads to a diagonal matrix with elements $c_m(t,t') \equiv \sqrt{\Gamma}\sqrt{(m+1)(N-m)}\xi_m(t')$. The trace of that matrix, $\sum_m c_m(t,t')$ yields the desired correlation function (6).

Numerical evaluation of $c_m(t,t')$ is possible for quite large values of $N$ due to the simple initial condition and the special properties of the time evolution, ensuring that the states and operators can be represented by vectors of length $N+1$ and $N$ instead of full matrices of size $N+1 \times N+1$. The first few modes $v_i(t)$ are plotted in fig. 1 for $N = 60$. The shape of the modes (in suitably scaled units) do not change appreciably for larger values of $N$, and an approximately constant fraction, $n_1 = 90.2\%$, of the emitted quanta are found in the dominant mode for all $N > 60$.

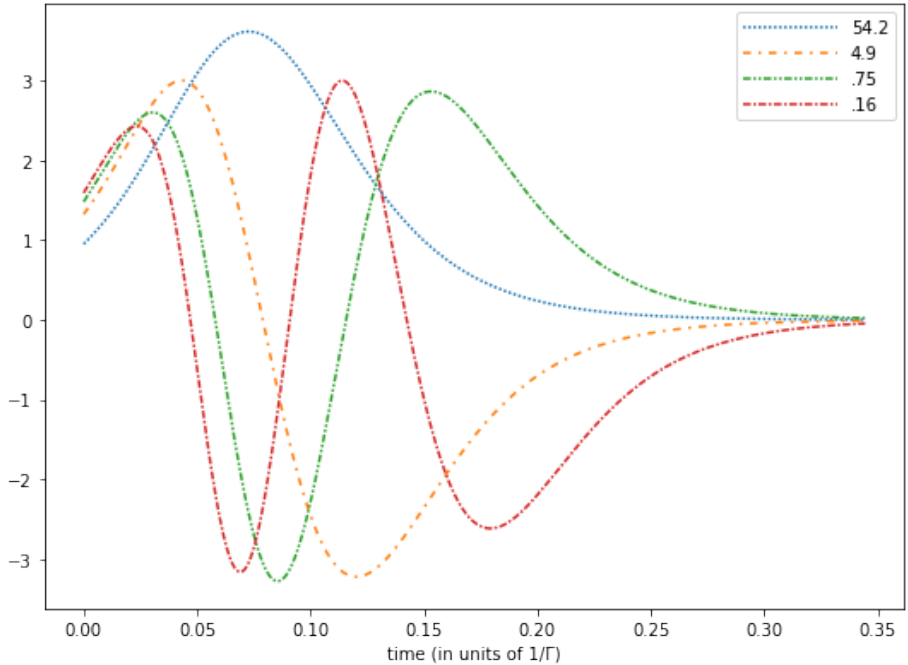

**Figure 1.** The normalized field modes occupied by superradiant emission by 60 atoms. The horizontal axis denotes time in units of $\Gamma^{-1}$, and the different modes are labelled by their occupations, which to numerical error sum to 60. The modes are found by solution of the master equation and use of the quantum regression theorem, followed by a numerical diagonalization of the two-time correlation function.

## 3. Analytical Approximation

In this section, we make a series of approximations to obtain analytic expressions for the field correlation function and the superradiant field modes emitted by a large number of atoms $N$. In the large $N$ limit, we can simplify some expressions by setting $\Gamma_m \simeq \Gamma_{m+1} \simeq \Gamma m(N-m)$ in which case the off diagonal elements $\xi_m$ in (8) solve the same equations as the diagonal density matrix elements $\pi_m$,

$$\dot{\xi}_m \approx -\Gamma_m \xi_m + \Gamma_{m+1} \xi_{m+1} \qquad (9)$$

We shall also treat $m$ as a continuous variable and the variation of different functions of $m$ by continuous derivatives. When we approximate $m$ as a continuous variable running from 0 to $N$, both $\pi_m$ and $\xi_m$ can be treated as functions of time and the continuous argument $m$ on the domain $(0, N)$. This may not be a good Ansatz for the initial value of $m = N$, but for now we shall think of the initial condition $\pi(t = 0, m)$ as a smooth function that peaks sharply near $m = N$.

Approximating differences between discrete elements by a first order partial derivative expansion

$$h(m+1) - h(m) \simeq \frac{\partial}{\partial m} h(m) \cdot 1 \tag{10}$$

the dynamical equations (5,9) for $g = \pi$ or $\xi$ can be written as

$$\dot{g}(t,m) = \frac{\partial}{\partial m}(\Gamma(m)g(t,m)). \tag{11}$$

This equation is of first order in both $t$ and $m$ and is solved by the Ansatz

$$g(t,m) = \frac{f(T(m) + Nt)}{\Gamma(m)} \tag{12}$$

where

$$T(m) = N \int_0^m \frac{1}{\Gamma(m')} dm' = \frac{1}{\Gamma} \log(\frac{m}{N-m}), \tag{13}$$

and the function $f$ is uniquely determined by the initial condition,

$$\pi(0,m) = \frac{f(T(m))}{\Gamma(m)}. \tag{14}$$

Using Eq.(12), we obtain the time evolved state by a scaling factor and a shift of the argument,

$$\pi(t,m) = f(T(m) + Nt)/\Gamma(m) = \frac{\Gamma(m')}{\Gamma(m)} \pi(0, m') \tag{15}$$

where $m'$ is found by solving the equation, $T(m) + Nt = T(m')$. At this point, the emitted field intensity can be calculated as weighted sum (integral) of decay rates, $\int \pi(t,m)\Gamma(m)dm$, but it will also follow from our correlation function analysis (see (24) and (25)).

At the time $t$ we define $\xi(t,m) = \sqrt{\Gamma(m)}\pi(t,m)$ and we use this as initial condition for Eq.(11). The solution at time $t'$ is given as above: $\xi(t',m) = \frac{\Gamma(m')}{\Gamma(m)} \xi(t,m')$, where $m'$ is the solution of the equation, $T(m) + Nt' = T(m') + Nt$. Finally, we represent the action by $\sqrt{\Gamma} J^+(t')$ to obtain the diagonal elements, $c(t',t,m) = \sqrt{\Gamma(m)} \xi(t',m)$.

For ease of notation, we define the rescaled time arguments $\tau$, $\tau'$ to represent $N\Gamma t$, $N\Gamma t'$, and we use (13) and its inverse $m(T) = N\frac{e^{\Gamma T}}{1+e^{\Gamma T}}$ as well as the expression $\Gamma(T) = \Gamma N^2 \frac{e^{\Gamma T}}{(1+e^{\Gamma T})^2}$ for $\Gamma(m)$ in terms of $T(m)$. With these relations it is possible to obtain the values of the arguments $m'$ in (15) as function of $t$ and in the similar expression for $\xi(t,m)$ as function of $t' - t$.

In order to determine the function $f$, we assume an initial atomic density matrix with diagonal elements , leading to the (unnormalized) continuous representation

$$\rho_{mm}(0) \to \pi(0,m) = e^{-\lambda(N-m)}. \tag{16}$$

The inital state is prepared with $m = N$, but we shall assume a finite value of $\lambda$ to compensate for the fact that the rate $\Gamma(m)$ in our our continuous description vanishes and does not yield the correct discrete decay rate when $m = N$. We find that $\lambda = 0.96$ yields good agreement between our analytical solutions and the exact numerical simulations for finite $N$, as well as with a simple mean field approximation for large $N$. Our results are not sensitive to small changes in $\lambda$ and we set $\lambda$ to a fixed value of 0.96 throughout.

From the initial condition (16) we can deduce $f$ via (14),

$$f(T(m)) = \Gamma m(N-m)e^{-\lambda(N-m)} \tag{17}$$

and evaluating the different expression, we finally obtain

$$\begin{aligned} c(\tau', \tau, m) &= f\Big(\frac{1}{\Gamma}\log\big(\frac{m}{N-m}\big) + Nt'\Big)\frac{e^{(\tau'-\tau)/2}}{N + m(e^{\tau'-\tau} - 1)} \\ &= e^{(3\tau'-\tau)/2}\frac{\Gamma m(N-m)e^{-\lambda N(N-m)/(N+m(e^{\tau'}-1))}}{(N + m(e^{\tau'-\tau} - 1))(N + m(e^{\tau'} - 1))^2}. \end{aligned} \tag{18}$$

The remaining trace of the operator product is calculated by summing (integrating) this expression over $m$, and to this end, we employ the substitution $y = m/N$ to obtain the definite integral representation of the field correlation function,

$$\langle J^+(\tau')J^-(\tau)\rangle \propto e^{(3\tau'-\tau)/2}\int_0^1 \frac{y(1-y)e^{-\lambda N(1-y)/(1+y(e^{\tau'}-1))}}{(1+y(e^{\tau'-\tau}-1))(1+y(e^{\tau'}-1))^2}dy \tag{19}$$

Note that this expression is not normalized, but we can renormalize the final integrated intensity to the number of excited atoms $N$.

The maximum of the correlation function is near the diagonal $\tau = \tau'$ and well separated from the origin $\tau = \tau' = 0$. Thus. we can apply $e^{\tau'} >> e^{\tau'-\tau}$ and $e^{\tau'}-1 \approx e^{\tau'}$, and even, $1 + ye^{\tau'} \approx ye^{\tau'}$, to simplify calculations. We have thus verified that the approximation

$$\langle J^+(\tau')J^-(\tau)\rangle \propto e^{-(\tau'-\tau)/2}\int_0^1 \frac{1-y}{y(1+y(e^{\tau'-\tau}-1))}e^{-\lambda N e^{-\tau'}(1-y)/y}dy \tag{20}$$

applies for large $N$. In terms of the well studied exponential integrals,

$$ei(x) = -\int_{-x}^\infty \frac{e^{-t}}{t}dt \tag{21}$$

we obtain,

$$\langle J^+(t')J^-(t)\rangle \propto \frac{\exp[\lambda N e^{-\tau'} - \frac{1}{2}(\tau+\tau')]}{e^{\tau'-\tau}-1}\bigg(ei(-\lambda N e^{-\tau'}) - \exp[\tau'-\tau+\lambda N(e^{-\tau}-e^{-\tau'})]ei(-\lambda N e^{-\tau})\bigg) \tag{22}$$

Finally, the function $ei(x)$ can be approximated by $ei(-x) \approx -(1/2)e^{-x}\log(1+2/x)$,

leading to a much simpler expression for the correlation function:

$$\langle J^+(x')J^-(x)\rangle = \frac{\sqrt{xx'}}{x'-x}\big(s(x')-s(x)\big) \tag{23}$$

where $x = e^\tau$, $x' = e^{\tau'}$ and $s(x) = \log(1+\beta x)/x$ with $\beta = 2/(\lambda N)$.

The equations (22) and (23) are the main results of this analysis, and in the following they will be used to identify the most populated field eigenmodes. Note that they also yield approximate expressions for the intensity when $t' = t$ ($\tau = \tau'$) where they can be evaluated by continuity of the expression and l'Hôpital's rule.

Eq.(22) thus yields the analytic expression for the intensity

$$I(\tau) = I_0 e^{-\tau}\Big(1 + e^{\lambda N e^\tau}(1+\lambda N e^{-\tau})ei(-\lambda N e^{-\tau})\Big) \tag{24}$$

while Eq.(23) yields the simpler expression

$$I(\tau) = I_0\Big(\frac{\log(1+\beta e^\tau)}{e^\tau} - \frac{\beta}{1+\beta e^\tau}\Big) \tag{25}$$

We can compare the solutions with both the exact numerical simulation for small $N$ and with the mean field theory approximation for large $N$, which yields an intensity proportional to $\text{sech}^2(N(t-t_D)/2)$ where $t_D = \log(N)/(\Gamma N)$ [32]. Our Eq.(24) is indistinguishable from the exact solution for $N > 100$ while (25) matches the intensity profile well but not perfectly.

Our expressions for the correlation function can both be represented as matrices on a temporal grid and the modes can be found by their numerical diagonalization. By $N = 100$, the difference between our numerical diagonalization of the analytical expression (22) and the exact eigenmodes is only barely discernible by eye. If we use the approximation for the exponential integral, leading to (23), there is still reasonably good agreement, as can be seen in fig. 2. The normalized eigenvalues by our analytical treatment (23) ((22)) are found to be .905 (.904), .083 (.081), and .012 (.012), for the three most populated modes.

The analytic expressions for the correlations functions can be applied for values of $N$ well beyond what is tractable numerically by the master equation, and we can employ different efficient methods to find the most populated eigenmodes $v_i(t)$ according to Eq.(7). As we only expect our analysis to become more accurate for larger values of $N$, we have demonstrated a universal property of the mode content of superradiant emission in the Dicke model.

In the Appendix we derive a set of iterative first-order differential equations to find the eigenmodes even more efficiently than by numerical diagonalization of the two-time correlation function. The results are plotted in fig. 3 for $N = 10^8$. While the occupation numbers have shifted by a few percent, the overall picture is the same; a large majority but not all of the photons occupy the first mode, and each successive mode contains significantly fewer photons than the one before it.

## 4. Conclusion

We have revisited the Dicke model of superradiance, a small sample of two level atoms coupled to the electromagnetic vacuum and initially in the maximally excited state.

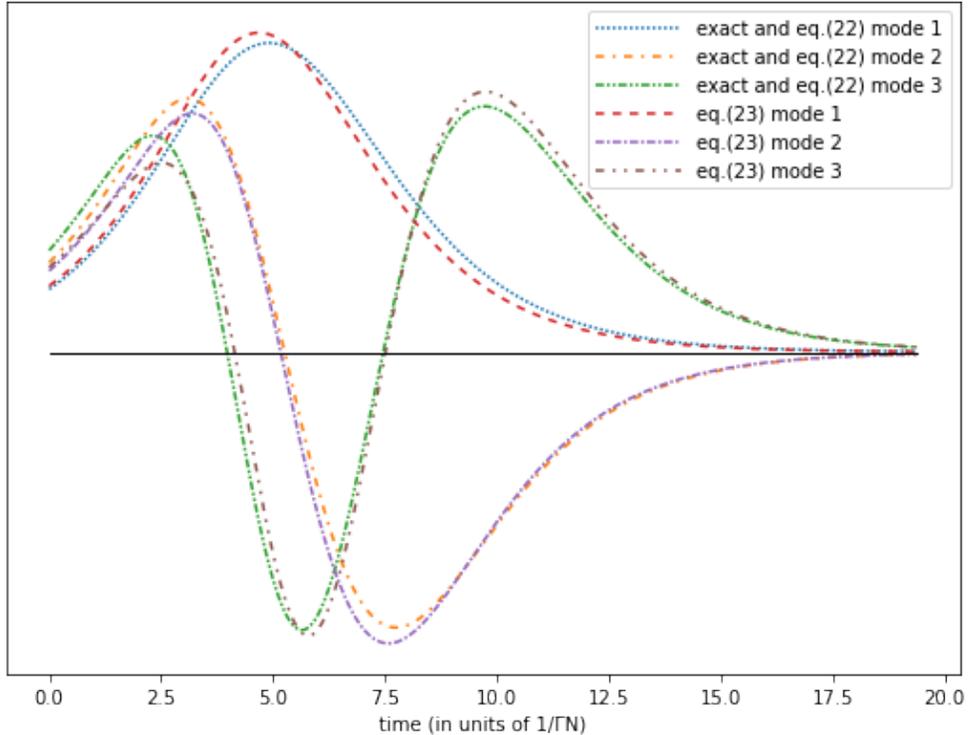

**Figure 2.** The first three eigenmodes for the exact numerical model, which overlaps exactly with the analytic model (22), and the approximate analytic model from (23), diagonalized numerically. Time is in units of $1/\Gamma N$ and $\lambda = .96$ and the vertical axis has an arbitrary scale (the zero line has been plotted in black). The exact (approximate) eigenvalues for mode 1, 2, and 3, normalized such that all eigenvalues sum to 1, are found to be .905 (.904), .083 (.081), and .012 (.012), respectively.

Using the master equation and quantum regression theorem, we evaluated the two-time correlation function of the field emitted by the atoms, and we found the dominant modes of outgoing photons and their relative occupations. Next, we derived excellent analytical approximations for the correlation function, and both the emitted intensity, the dominant modes and their relative occupations agreed well with the exact numerical solutions. Our method does not rely on symmetry breaking and a first order mean field theory, and it employs the quantum regression theorem to calculate two-time correlations in a quite straightforward manner, as long as the Hamiltonian and damping mechanisms do not induce coherences in the system. While we were able to obtain good analytical approximations for the case of Dicke superradiance, we note that the continuous description of the number of excitations in the system in the large $N$ limit would also permit purely numerical solutions and benefit the analysis of other problems without restriction to small values of $N$.

## 5. Acknowledgemnt

This work has been supported by the US ARL-CDQI program through cooperative agreement W911NF-15-2-0061, and the Danish National Research Foundation through the Center of Excellence "CCQ" (Grant agreement no.: DNRF156).

## 6. Appendix. Electric Field Mode Shapes

Remarkably, thanks to our lack of proper normalization which would be a function of $\beta$, the following useful relationship follows from (23)

$$\partial_\beta \langle J^+(z') J^-(z) \rangle = \frac{\sqrt{zz'}}{(1+z)(1+z')} \tag{26}$$

where $z = \beta x$ and $z' = \beta x'$.

In terms of the $z$ and $z'$ variables we can decompose the correlation function

$$\langle J^+(z') J^-(z) \rangle = \sum_i \lambda_i w_i(z) w_i^*(z'), \tag{27}$$

with values $\lambda_i$ and functions $w_i(z)$ which are not orthogonal and are therefore not uniquely determined by (27). However we will be able to extract a small set of such functions and use them to construct the most populated orthogonal field eigenmodes.

We will require that the $\lambda_i$ do not depend on $\beta$, and that the functions $w_i$ may not be normalized. From (26), we can infer that $w_i(z) = \sqrt{\beta} u_i(z)$ for some $u_i(y)$ which does not depend explicitly on $\beta$. Using this observation, along with the fact that $\partial_\beta z = x$, we take the derivative of the correlation function with respect to $\beta$ and obtain

$$\begin{aligned}\frac{\sqrt{zz'}}{(1+z)(1+z')} &= \sum_i \lambda_i \left( u_i(z) u_i^*(z') + \beta \Big( u_i'(z) x u_i^*(z') + u_i(z) u_i^{*\prime}(z') x' \Big) \right) \\ &= \sum_i \lambda_i \left( u_i(z) u_i^*(z') + z u_i^{*\prime}(z) u_i(z') + z' u_i(z) u_i^{*\prime}(z') \right)\end{aligned} \tag{28}$$

Re-arranging the terms, this can be written as

$$\frac{\sqrt{zz'}}{(1+z)(1+z')} = \sum_i \lambda_i \left( (u_i(z) + z u_i'(z))(u_i^*(z') + z' u_i^{*\prime}(z')) - zz' u_i'(z) u_i^{*\prime}(z') \right) \tag{29}$$

In our case the two-time correlation function is real, and therefore $u_i$, $w_i$, and $\lambda_i$ can also be chosen real, but we retain the complex notation for the potential application to other problems with complex correlation functions.

We can now see that the equality will be satisfied if

$$\begin{aligned}\sqrt{\lambda_1}\big(u_1(z) + z u_1'(z)\big) &= \sqrt{z}/(1+z) \\ \sqrt{\lambda_i} z u_i'(z) &= \sqrt{\lambda_{i+1}}\big(u_{i+1}(z) + z u_{i+1}'(z)\big)\end{aligned} \tag{30}$$

The left term in the sum with $i = 1$ gives the left hand side of the equation, and the right term in the sum for each $i$ cancels with the left term from the next $i$. If we truncate the sum at some finite $i$ then there will always be a remainder, but as long as $\lambda_i |u_i|^2$ shrinks to zero as $i$ grows, as we will find to be the case, this remainder can be approximately ignored.

Let us define $u_i(z) = c_i g_i(z)$ where the $g_i$ obey the simpler relations

$$g_1(z) + zg_1'(z) = \sqrt{z}/(1+z)$$
$$zg_i'(z) = g_{i+1}(z) + zg_{i+1}'(z) \tag{31}$$

The $\lambda_i c_i^2$ must all equal one another in order for the cancellation of successive terms in the sum of the right hand side of (29) to be complete. Additionally, in order for the left hand side of (29) to be generated by the term with $i = 1$, $\lambda_1 c_1^2$ must equal 1, and thus $\lambda_i c_i^2 = 1$ for all $i$.

The differential equations (31) can easily be integrated iteratively up to some finite $i$, and therefore the $w_i(z) = \sqrt{\beta/\lambda_i} g_i(z)$ can be identified.

While the functions identified indeed decompose the correlation functions, they do not form an orthonormal basis, when we consider them as function of the time argument $w_i(t) := w_i(z(t))$. But they form a useful starting point to identify the temporal eigenmodes. To that end we define the temporal overlap of any two functions

$$\langle w_j, w_i \rangle = \int_0^\infty w_j^*(t) w_i(t) dt \tag{32}$$

and we assume that a temporal mode $\psi$ can be expanded as

$$\psi = \sum_i a_i w_i \tag{33}$$

If $\psi$ is an eigenmode of the superradiant output field with the photon occupation $\nu$ temporal, it is an eigenfunction of an integral equation with the correlation function as integration kernel,

$$\int_0^\infty \langle J^+(t') J^-(t) \rangle \psi(t) dt = \nu \psi(t') \tag{34}$$

I.e.,

$$\sum_i a_i \sum_j \lambda_j \int_0^\infty w_j(t') w_j^*(t) w_i(t) dt$$
$$= \sum_j w_j(t') \lambda_j \sum_i a_i \langle w_j, w_i \rangle = \nu \sum_j a_j w_j(t') \tag{35}$$

from which we can deduce that

$$\nu a_j / \lambda_j = \sum_i a_i \langle w_j, w_i \rangle \tag{36}$$

This is a generalized eigenvalue problem, $A\vec{a} = \nu B\vec{a}$ with $A_{ij} = \langle w_i, w_j \rangle$ and $B_{ij} = \delta_{ij} \lambda_j^{-1}$. As after, e.g., the fifth eigenvalue, the occupations become much less than .01%, we may evaluate just a few functions $w_i$ and solve the corresponding low dimensional eigenvalue problem. This yields the different eigenmodes $v_i$ and their photon number populations.

Using the method described above, we can plot the eigenmodes for extremely large $N$. The results for $N = 10^8$ are displayed in fig. 3.

In the approximation (23) the enormously large prefactors cancel the extremely small value of the exponential integrals to form an expression that is readily calculated even for very large $N$. While we have found that the analytical expression (22) is almost exact for large $N$, it is not straightforwardly evauated due to the combination of exponentially large and small arguments. To make use of (22), one may thus have recourse to some of the more elaborate approximate expressions for the exponential integral, see , e.g., [33].

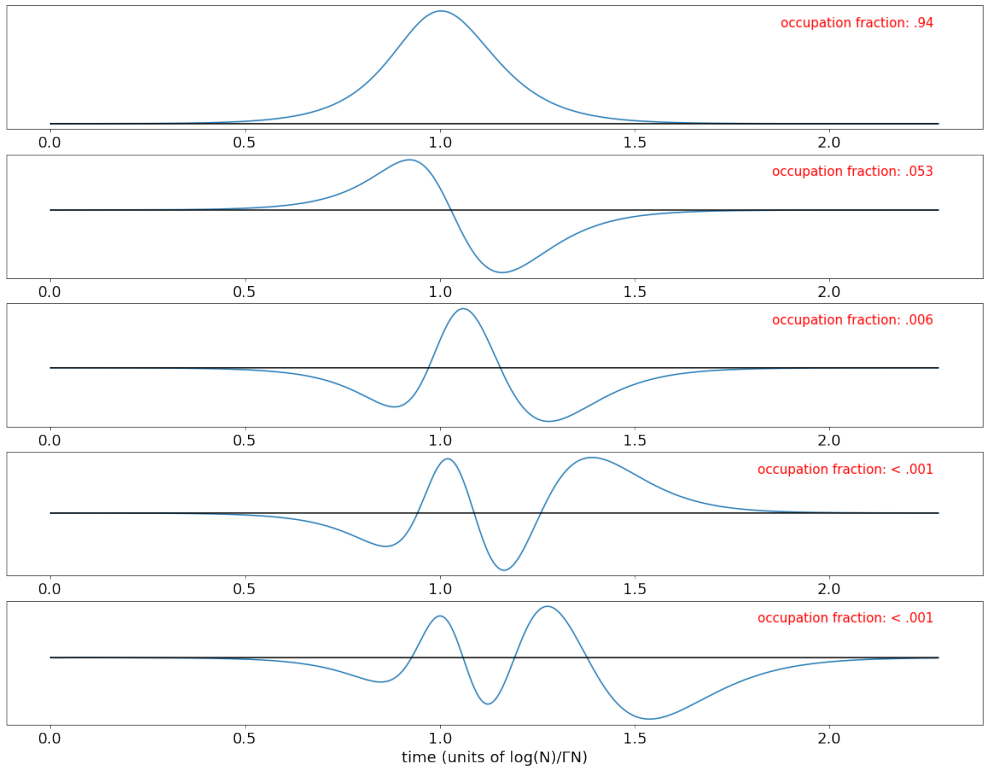

**Figure 3.** The first five eigenmodes of the two-time correlation function, according to (23) with $\lambda = .96$ and $N = 10^8$ and calculated via the process outlined in the appendix. Time is in units of $\log(N)/\Gamma N$ along the horizontal axis, while the vertical variation of the mode functions is shown in arbitrary units.

Associated with these modes we have normalized eigenvalues, representing the fraction of photons occupying that modes, equal to .94, .053, .0060, .00092, and .000055. This is qualitatively similar to the distribution of eigenvalues in the more exact, small $N$ cases studied above. The differences can be ascribed to the approximation of the exponential integral, as well as the different value of $N$. It is worth noting that even though $N$ has changed by 6 orders of magnitude from the $N = 100$ case, the occupation of the dominant eigenmodes has only changed by a few percent.

## References

[1] R. Dicke, *Coherence in Spontaneous Radiation Processes.* Phys.Rev. 93 99– 110 (1954)

[2] D. Meiser, Jun Ye, D. R. Carlson, and M. J. Holland, *Prospects for a mHz-linewidth laser.* Phys. Rev. Lett. 102, 163601 (2009)

[3] A. A. Svidzinsky, L. Yuan, M. O. Scully, *Quantum Amplification by Superradiant Emission of Radiation*, Phys. Rev. X, Volume 3, Issue 4 (2013)

[4] S. Inouye, A. P. Chikkatur, D. M. Stamper-Kurn, J. Stenger, D. E. Pritchard, W. Ketterle, *Superradiant Rayleigh scattering from a Bose-Einstein condensate.* Science 285, 571–574 (1999)

[5] A. Walther, A. Almari, S. Kröll, *Experimental superradiance and slow-light effects for quantum memories*, Phys. Rev. A 80 (2009)

[6] E. Novais, E. R. Mucciolo, and H. U. Baranger, *Resilient Quantum Computation in Correlated Environments: A Quantum Phase Transition Perspective*, Phys. Rev. Lett. **98**, 040501 (2007).

[7] B. Lemberger, D. D. Yavuz, *Effect of correlated decay on fault-tolerant quantum computation*, Phys. Rev. A **96**, 062337 (2017).

[8] V. Paulisch, M. Perarnau-Llobet, A. González-Tudela, and J. I. Cirac, *Quantum metrology with one-dimensional superradiant photonic states.* Phys. Rev. A 99, 043807 (2019)

[9] M. Gross, S. Haroche, *Superradiance: An essay on the theory of collective spontaneous emission.* Physics Reports, Volume 93, Issue 5, p. 301-396 (1982)

[10] R. Reimann, W. Alt, T. Kampschulte, T. Macha, L. Ratschbacher, N. Thau, S. Yoon, D. Meschede, *Cavity-modified collective Rayleigh scattering of two atoms.* Phys. Rev. Lett. 114, 023601 (2015)

[11] I. E. Mazets and G. Kurizki, *Multi-atom coooperative Emission Following Single Photon Absorption: Dicke-State Dynamics*, J. Phys. B: At. Mol. Opt. Phys. **40**, F105 (2007).

[12] Aleksandrov, A.I., Aleksandrov, I.A., Zezin, S.B. et al. *Radio-frequency superradiance induced by the rheological explosion of polymer composites containing paramagnetic cobalt complexes*, Russ. J. Phys. Chem. B 10, 69–76 (2016)

[13] A. Angerer, K. Streltsov, T. Astner, S. Putz, H. Sumiya, S. Onoda, J. Isoya, W. Munro, K. Nemoto, J. Schmiedmayer, J. Majer, *Superradiant emission from colour centres in diamond*, Nature Physics. 14. 10.1038 (2018)

[14] M. Scheibner, T. Schmidt, L. Worschech, A. Forchel, G. Bacher, T. Passow, D. Hommel, *Superradiance of quantum dots.* Nat. Phys. 3, 106–110 (2007)

[15] V. E. Elfving, S. K. Das, and A. S. Sørensen, *Enhancing quantum transduction via long-range waveguide-mediated interactions between quantum emitters*, Physical Review A, 100(5) (2019)

[16] A. Asenjo-Garcia, M. Moreno-Cardoner, A. Albrecht, H.J. Kimble, and D.E. Chang, *Exponential Improvement in Photon Storage Fidelities Using Subradiance and "Selective Radiance" in Atomic Arrays*, Phys. Rev. X 7, 031024 (2017)

[17] A. Goban, C. L. Hung, J.D. Hood, S. P. Yu, J.A. Muniz, O. Painter, and H.J. Kimble, *Superradiance for Atoms Trapped along a Photonic Crystal Waveguide*, Phys. Rev. Lett. 115, 063601 (2015)

[18] A. Albrecht, L. Henriet, A. Asenjo-Garcia1, P. B. Dieterle, O. Painter, and D. E. Chang, *Subradiant states of quantum bits coupled to a one-dimensional waveguide*, New Journal of Physics, Volume 21, (2019)

[19] W. Guerin, M. O. Araújo, R. Kaiser, *Subradiance in a Large Cloud of Cold Atoms*, Phys. Rev. Lett. 116 (2016)

[20] R. Glauber, F. Haake, *The initiation of superfluorescence*, PHys. Lett. A 68 (1978)

[21] F. Haake, H. King, G. Schröder, J. Haus, R. Glauber, *Fluctuations in superfluorescence*, Phys. Rev. A 20 2047 (1979)

[22] H. J. Kimble, L. Mandel, *Theory of resonance fluorescence*, Phys. Rev. A 13, 6 (1976)

[23] U. M. Titulaer, R. J. Glauber, *Density Operators for Coherent Fields*, Phys. Rev. 145 1041 (1966)

[24] P. M. G. Raymer, Z. W. Li, I. A. Walmsley, *Temporal Quantum Fluctuations in Stimulated Raman Scattering: Coherent-Modes Description*, Phys. Rev. Lett. 63, 1586 (1989)

[25] M. G. Raymer, I. A. Walmsley, *Temporal modes in quantum optics: then and now*, Phys.

Scr. 95 (2020)

[26] B. Brecht, Dileep V. Reddy, C. Silberhorn, M. G. Raymer, *Photon Temporal Modes: A Complete Framework for Quantum Information Science*, Phys. Rev. X 5, 041017 (2015)
[27] C. Fabre, N. Treps, *Modes and states in quantum optics*, arXiv:1912.09321 (2019)
[28] C. Gardiner, P. Zoller, *Quantum noise: a handbook of Markovian and non-Markovian quantum stochastic methods with applications to quantum optics*, SPringer Science & Business Media (2004)
[29] D. Kleppner, *Inhibited Spontaneous Emission*, Phys. Rev. Lett. 47, 4 (1981)
[30] J. Mercer, *XVI. Functions of positive and negative type, and their connection the theory of integral equations*, Philosophical Transcations of the Royal Society A, 209 (1909)
[31] B. R. Mollow, *Power Spectrum of Light Scattered by Two-Level Systems*, Phys. Rev. 188 (1969)
[32] Benedict, M. (Ed.), *Super-radiance*, Boca Raton: CRC Press (1996)
[33] D. A. Barry, J. -Y. Parlange, L. Li, *Approximation for the exponential integral (Theis well function)*, Journal of Hydrology 227 (2000)